\documentclass[letterpaper]{article} 
\usepackage{aaai2026}  
\usepackage{times}  
\usepackage{helvet}  
\usepackage{courier}  
\usepackage[hyphens]{url}  
\usepackage{graphicx} 
\usepackage{natbib}  
\usepackage{caption} 
\usepackage{algorithm}
\usepackage{algorithmic}
\usepackage{enumitem} 

\usepackage{newfloat}
\usepackage{listings}
\DeclareCaptionStyle{ruled}{labelfont=normalfont,labelsep=colon,strut=off} 
\floatstyle{ruled}
\newfloat{listing}{tb}{lst}{}
\floatname{listing}{Listing}
\title{STEM Faculty Perspectives on Generative AI in Higher Education}
\author{
   Akila de Silva\textsuperscript\equalcontrib,
    Isabel Hyo Jung Song \textsuperscript\equalcontrib,
    Hui Yang\textsuperscript\equalcontrib,
    Shah Rukh Humayoun\textsuperscript\equalcontrib
}
\affiliations{
    Department of Computer Science\\
    San Francisco State University\\
    1600 Holloway Ave\\
    San Francisco, CA 94132 USA\\
    \{desilva, hjsong, huiyang, humayoun\}@sfsu.edu\\
}

\usepackage{bibentry}

\begin{document}

\maketitle

\begin{abstract}
Generative artificial intelligence (GenAI) tools are increasingly present in higher education, yet their adoption has been largely student-driven, requiring instructors to respond to technologies already embedded in classroom practices. While some faculty have embraced GenAI for pedagogical purposes such as content generation, assessment support, and curriculum design, others approach these tools with caution, citing concerns about student learning, assessment validity, and academic integrity. Understanding faculty perspectives is therefore essential for informing effective pedagogical strategies and institutional policies. In this paper, we present findings from a focus group study with 29 STEM faculty members at a large public university in the United States. We examine how faculty integrate GenAI into their courses, the benefits and challenges they perceive for student learning, and the institutional support they identify as necessary for effective and responsible adoption. Our findings highlight key patterns in how STEM faculty engage with GenAI, reflecting both active adoption and cautious use. Faculty described a range of pedagogical applications alongside concerns about student learning, assessment, and academic integrity. Overall, the results suggest that effective integration of GenAI in higher education requires rethinking assessment, pedagogy, and institutional governance in addition to technical adoption.
\end{abstract}

%

\section{Introduction}

Generative Artificial Intelligence (GenAI) 
refers to a type of artificial intelligence systems that can create new content, such as images, text, music, or videos, based on patterns learned from large datasets. By utilizing neural networks, particularly transformers~\cite{vaswani2017attention}, GenAI 
systems can generate realistic outputs in response to prompts and instructions. 
GenAI has been applied to many application domains, for example, automated software code generation~\cite{odeh2024comparative}, synthesis of realistic images for digital media~\cite{elasri2022image}, and summarizing technical content~\cite{guan2020survey}.


While industries have utilized GenAI strategically, its use in higher education has been largely student-driven, compelling instructors to respond to a technology that is already widely available in the classroom. As a result, some instructors have embraced the use of GenAI, while others have expressed skepticism and concern. Those who embrace GenAI have incorporated it in a variety of pedagogical purposes. These include its use as a grading tool~\cite{tobler2024smart}, for curricular design~\cite{ullmann2024towards}, for course content generation~\cite{owen2025beyond}, and for personalized education~\cite{udeh2025role}. To further support these efforts, some educational institutions have made premium subscriptions to GenAI tools freely available to students and faculty, signaling a broader institutional commitment to embrace GenAI tools for teaching and learning~\cite{barajas2025inside, ASU2025OpenAIExpansion}. 

In contrast to these adoption efforts, other faculty approach GenAI with caution, raising concerns about its impact on learning, assessment, and academic integrity. The foremost among these concerns are the erosion of academic integrity and deskilling fundamental skills such as critical thinking and problem solving. The widespread availability of GenAI tools capable of producing sophisticated text is expected to increase undetectable forms of plagiarism~\cite{gallent2023impact, eke2023chatgpt, bittle2025generative}.  Faculty concerns also extend to the trustworthiness of GenAI outputs. Large language models have a documented tendency to \textit{hallucinate}, producing plausible but incorrect information. This presents a substantial pedagogical risk, especially for students who lack the domain knowledge required to critically assess such outputs~\cite{kamel2024understanding, alimardani2024human}. Additionally, faculty are concerned with the ethical and equity implications associated with GenAI. These tools can perpetuate the biases inherent in the training data~\cite{garcia2025ethical, quince2024current}.

In light of the diverse faculty responses to GenAI use in the classroom, ranging from active adoption to skepticism, a clear understanding of faculty perspectives is a necessary precursor to the development of effective institutional policies and pedagogical strategies. This study draws on focus group discussions with $29$ faculty members in the College of Science and Engineering (CoSE) at San Francisco State University (SFSU), a large public and Hispanic-serving institution that enrolls a substantial portion of Pell grant eligible students. As an access-oriented university serving many first generation and historically underrepresented students, SFSU provides an important context for examining faculty's perspectives on generative AI use in higher education.   


To examine these issues, the focus group discussions were designed around three core research questions:  
\begin{enumerate}[label=RQ\arabic*, leftmargin=*, align=left]
    \item How do faculty integrate GenAI into course design and learning activities? 
    \item What benefits and challenges have instructors observed regarding student                           learning? 
    \item What institutional resources and policies are needed to support effective                GenAI adoption? 
\end{enumerate}

\noindent The focus group discussions gathered qualitative and quantitative data on faculty perceptions, practices, and institutional needs on the use of GenAI. Our findings highlight several key patterns in how faculty engage with GenAI in STEM education. First, rather than reducing faculty workload, GenAI shifts instructional labor from content creation to expert curation, requiring faculty to review, refine, and verify AI-generated outputs. Second, while GenAI use is associated with higher assignment submission rates, faculty observed that it can mask gaps in students’ conceptual understanding, raising concerns about the validity of existing assessment practices. In response, faculty are experimenting with a dual strategy: reverting to more controlled evaluation formats (e.g., in-class exams and oral assessments) while also designing assignments that promote critical engagement with AI-generated outputs. Finally, our findings highlight the need for institutional approaches that balance university-wide policy guardrails with department-level autonomy, alongside sustained investments in faculty development and pedagogical support. Collectively, these results underscore that effective integration of GenAI in higher education requires not only technical adoption, but also rethinking assessment, pedagogy, and institutional governance under ongoing uncertainty.


\medskip
\noindent This paper makes the following contributions: 
\begin{itemize}
       \item Gains insights into faculty perspectives on GenAI use in higher education from a large public Hispanic-serving and access-oriented university.
    \item Identifies pedagogical benefits and challenges on using GenAI.
    \item Proposes faculty informed recommendations on institutional support. 
\end{itemize}
    


\section{Related Work   }

\subsection{Generative AI}
The literature characterizes GenAI as a class of artificial intelligence (AI) systems capable of producing novel content — including text, images, music, and video — by learning statistical patterns from large-scale datasets. Prior work notes that contemporary GenAI systems are predominantly built on neural network architectures, particularly transformer models~\cite{vaswani2017attention}, which enable the generation of coherent and contextually relevant outputs in response to user prompts. Research across multiple domains has documented the application of GenAI for tasks such as automated software code generation~\cite{odeh2024comparative}, the synthesis of realistic visual content for digital media~\cite{elasri2022image}, and the summarization of complex or technical information~\cite{guan2020survey}. As these capabilities continue to mature, scholars have increasingly examined how GenAI is being adopted, interpreted, and contested within higher education settings.

\subsection{Uses of GenAI in Higher Education}

A growing body of research documents how faculty have begun adopting GenAI tools across a range of instructional contexts. Prior studies report the use of GenAI for assessment-related tasks such as grading and feedback~\cite{tobler2024smart}, as well as for curriculum and instructional design~\cite{ullmann2024towards}, course content creation~\cite{owen2025beyond}, and personalized or adaptive learning experiences~\cite{udeh2025role}. Beyond individual faculty initiatives, institutional support has also emerged, with some higher education institutions providing faculty and students with access to premium GenAI tools. These efforts reflect a broader organizational commitment to integrating GenAI into teaching and learning practices at scale~\cite{barajas2025inside, ASU2025OpenAIExpansion}.However, alongside these emerging uses, researchers note that faculty adoption of GenAI is frequently accompanied by significant reservations regarding its educational implications.

\subsection{Concerns of GenAI in Higher Education}

Despite the growing adoption of GenAI in higher education, a substantial body of literature highlights significant concerns regarding its impact on teaching and learning. Chief among these are issues of academic integrity~\cite{evangelista2025ensuring}, including plagiarism, and the limitations of existing detection tools in reliably identifying AI-generated content~\cite{gallent2023impact, eke2023chatgpt, bittle2025generative}. Faculty have also expressed apprehension that widespread use of GenAI may undermine the development of core skills such as critical thinking and problem-solving~\cite{larson2024critical, helal2025impact}. Additional concerns relate to ethical and institutional challenges, including data privacy, algorithmic bias, and the lack of clear governance frameworks to guide responsible use~\cite{luckett2023regulating, raza2025responsible}. 

Despite increasing scholarly attention, existing work offers limited insight into how faculty, particularly in STEM disciplines, collectively interpret and negotiate the role of GenAI in higher education.

\section{Methods}

We designed a qualitative study using focus groups to investigate STEM faculty perspectives on the use of GenAI in teaching and learning. Focus groups were chosen to elicit shared practices, collective concerns, and cross-disciplinary discussion.

\subsection{Participants and Recruitment}
Participants were 29 faculty members from the College of Science and Engineering (CoSE) at San Francisco State University (SFSU), including lecturers and tenured/tenure-track faculty. Participants were recruited during Summer and Fall 2025 semesters via email invitations to reach faculty within the Department of Computer Science (CS) and CoSE. Participation was voluntary.

The recruitment email described the purpose of the study, focus group study procedures, time commitment, compensation, and data handling practices. All participants were provided the form of informed consent electronically via DocuSign prior of participation. The study protocol received Institutional Review Board (IRB) exemption approval in Summer 2025. Participants received a \$75 gift card as a token of appreciation.

\subsection{Focus Group Sessions}
We conducted seven focus group sessions, each lasting approximately 90 minutes. Sessions were held remotely via Zoom to maximize accessibility and participation across departments. Group sizes ranged from three to five participants, consistent with recommendations for fostering in-depth discussion while ensuring equitable participation.

Three sessions included only computer science faculty (three participants per session; nine participants total), allowing for discipline-specific discussion of GenAI practices. The remaining four sessions were interdisciplinary, including faculty from multiple CoSE departments; two of these sessions included computer science faculty alongside participants from other disciplines. This mixed-session design enabled comparison between discipline-specific and cross-disciplinary perspectives.

Each session was facilitated by at least two researchers, who moderated discussion, monitored participation, and ensured adherence to the study protocol. At the start of each session, a facilitator introduced the study goals and session structure, followed by brief participant self-introductions.

\subsection{Demographics of the Participants}

The 29 participants represented a range of departments within the College of Science and Engineering at SFSU. The largest group was from Computer Science (\(n = 11, 41.9\%\)), followed by the School of Engineering (\(n = 4, 12.9\%\)), Psychology (\(n = 4, 9.7\%\)), Mathematics (\(n = 3, 9.7\%\)), Chemistry and Biochemistry (\(n = 3, 9.7\%\)), the School of the Environment (\(n = 2, 6.5\%\)), Physics and Astronomy (\(n = 1, 3.2\%\)), and Biology (\(n = 1, 3.2\%\)). 
Among the CS faculty, four were lecturers and seven were tenured or tenure-track faculty. Among participants from the remaining STEM departments, three were lecturers and seventeen were tenured or tenure-track faculty.



\subsection{Research Questions}
The focus group protocol was informed by prior literature on AI in education and by institutional initiatives such as SFSU's Graduate Certificate in Ethical AI\footnote{https://bulletin.sfsu.edu/colleges/science-engineering/computer-science/certificate-ethical-artificial-intelligence/}. The study was guided by the following research questions:

\textbf{RQ1: How do faculty integrate generative AI into course design and learning activities?}  
To explore current instructional practices and perceived pedagogical value, participants were asked:
\begin{itemize}
    \item How do you integrate AI/GenAI tools into your course design and learning activities?
    \item What benefits have you observed from using AI/GenAI in your teaching?
\end{itemize}

\textbf{RQ2: What benefits and challenges do instructors observe regarding student learning?}  
This research question focused on faculty experiences, concerns, and ethical considerations related to student learning. Participants were asked:
\begin{itemize}
    \item What challenges or concerns have you encountered or anticipate regarding the use of AI/GenAI in education?
    \item What are your thoughts on the ethical implications of using AI/GenAI in education (e.g., plagiarism, bias, accessibility)?
    \item What curricular innovations do you think are necessary or beneficial in light of the rapid advancements in AI/GenAI?
\end{itemize}

\textbf{RQ3: What institutional resources and policies are needed to support effective generative AI adoption?}  
To identify institutional and structural needs, participants were asked:
\begin{itemize}
    \item What kinds of training or resources would help you learn more about AI/GenAI and use it effectively and ethically in your teaching?
    \item What policies or guidelines should the university implement regarding the ethical and responsible use of AI/GenAI in teaching and learning?
    \item What institutional or programmatic changes (e.g., departmental restructuring, resource allocation, new policies) would best support the effective integration of AI/GenAI into teaching and learning?
\end{itemize}

\subsection{Data Collection and Analysis}
At the beginning of each session, participants completed a short demographic questionnaire (six items) administered via the Qualtrics platform, 
the official web-based survey tool adopted at SFSU.
The focus group discussions then followed a semi-structured protocol consisting of eight open-ended questions organized around three research questions.

Discussions were conducted in a semi-controlled manner: while each participant was invited to respond to each question, facilitators encouraged follow-up questions, elaboration, and peer-to-peer dialogues to support a natural conversational flow. This approach enabled the research team to collect rich qualitative data while maintaining analytic focus.

All sessions were recorded using Zoom’s cloud recording feature, including audio, video, and automatically generated transcripts. In addition, researchers took detailed notes during the sessions to capture salient points, emergent themes, and contextual observations. All data were securely stored in a university-managed Box folder, with access restricted to the research team.

For data analysis, the automatically generated Zoom transcripts and AI-generated session summaries were first anonymized to remove personally identified information for each participant. Such data was then imported into Google NotebookLM\footnote{https://notebooklm.google/} to support exploratory qualitative analysis. NotebookLM was used to surface and organize prominent topics and discussion patterns across sessions, generating an interactive, hierarchical representation of themes discussed by participants (an example is shown in Figure~\ref{fig1}). The research team treated these AI-assisted outputs as analytic aids rather than definitive results. We cross-referenced and manually verified the themes identified by NotebookLM using the original transcripts 
and researchers' notes taken during these sessions.

\section{Results and Findings}

Based on responses to the closed-ended questionnaire completed at the beginning of the focus group sessions, participants reported substantial prior exposure to both artificial intelligence (AI) and generative AI (GenAI) technologies. With respect to AI broadly, most respondents indicated moderate to high familiarity: 37.9\% reported being very familiar (\(n=11\)) and 34.5\% reported being moderately familiar (\(n=10\)). An additional 24.1\% described themselves as slightly familiar (\(n=7\)), while a small number reported being extremely familiar (\(n=1, 3.4\%\)). No participants reported being unfamiliar with AI overall.

Familiarity with GenAI followed a similar but slightly more varied pattern. Most participants indicated at least moderate familiarity with GenAI, including those who were moderately familiar (40.0\%, \(n = 12\)) and very familiar (33.3\%, \(n = 10\)). Others reported being slightly familiar (20.0\%, \(n = 6\)), while very few participants indicated extreme familiarity (3.3\%, \(n = 1\)) or no familiarity at all (3.3\%, \(n = 1\)). Together, these results suggest that while the participant group was broadly knowledgeable about AI concepts, levels of experience with GenAI tools were more heterogeneous, providing a range of perspectives for the focus group discussions.  

Consistent with these familiarity levels, most participants reported some degree of engagement with GenAI in their teaching practices. A large majority indicated that they had used GenAI tools in an instructional context (\(n = 27\)), with only two participants reporting no prior use. Among those who reported using GenAI, faculty described a variety of instructional applications, including facilitating class discussions (\(n = 5\)), generating assessment or practice questions (\(n = 4\)), providing feedback to students (\(n = 2\)), supporting student research activities (\(n = 1\)), and automating aspects of grading (\(n = 1\)). Reported frequency of use varied across participants: GenAI was most commonly used occasionally (\(n = 12\)) or frequently (\(n = 9\)), with fewer participants reporting rare use (\(n = 6\)) or no use (\(n = 2\)). Together, these results indicate that while GenAI adoption among STEM faculty is already underway, patterns of use and levels of integration vary considerably, providing important context for the qualitative findings reported in the following sections.

\begin{figure}[t]
\centering
\includegraphics[width=0.9\columnwidth]{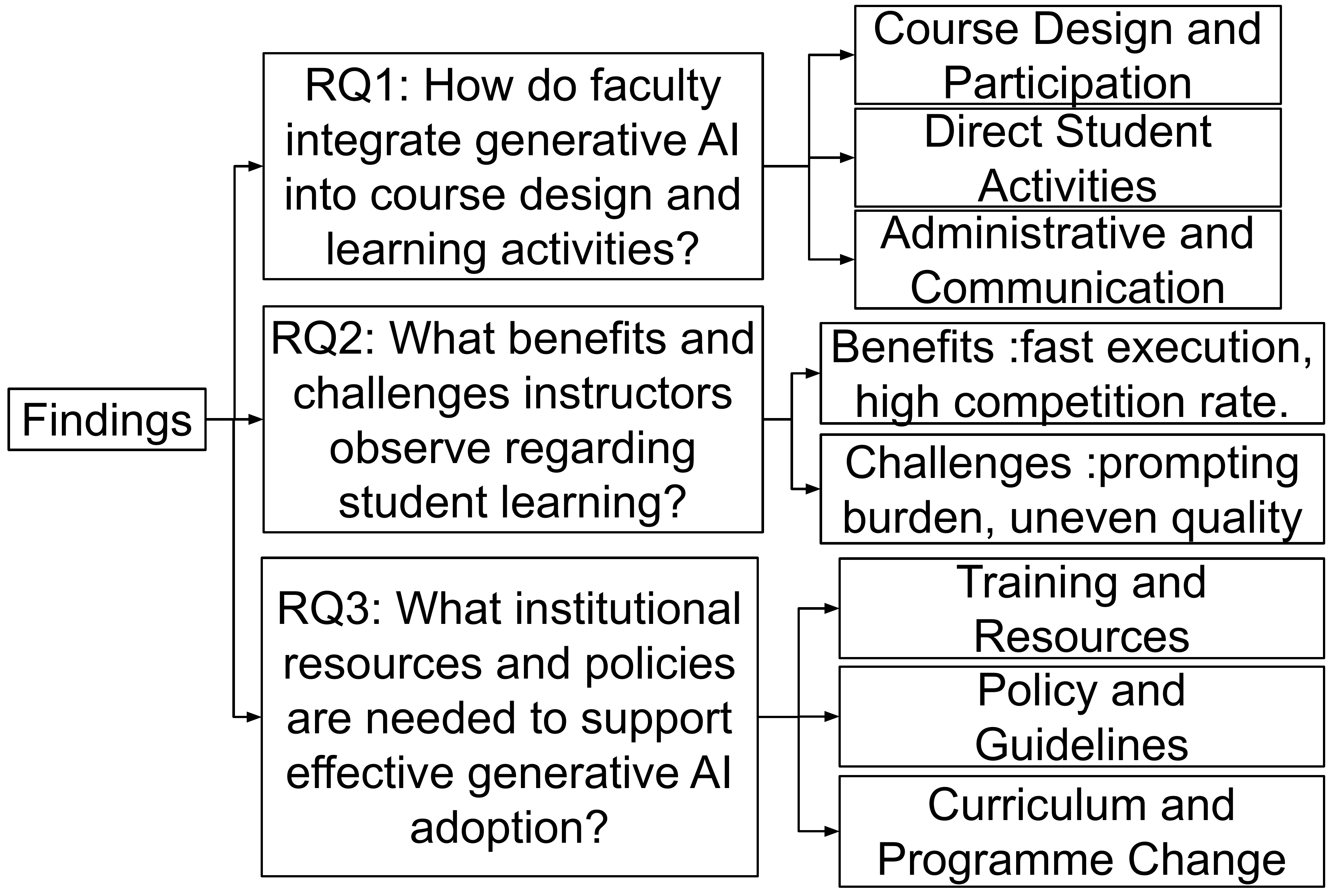} 
\caption{An Abbreviated Taxonomy of Our Findings}
\label{fig1}
\end{figure}

We next report findings aligned with the three research questions, examining how faculty integrate GenAI into course design and learning activities (RQ1), the benefits and challenges they observe for student learning (RQ2), and the institutional resources and policies needed to support effective adoption (RQ3) as shown in Figure~\ref{fig1}.

\subsection{RQ1: How do faculty integrate generative AI into course design and learning activities? }\label{RQ1}

Among the 29 faculty participants, 27 or $93\%$ reported using GenAI tools to facilitate both teaching and learning activities. Their use cases can be categorized into the following three categories: 1) facilitating course design and preparation, 2) directly supporting students' learning, and 3) assisting with administrative tasks and communications.

\subsubsection{\textit{Facilitating course design and preparation:}} As mentioned above, a large majority of the faculty participants shared that they used generative AI tools for course design and preparation. The most frequently cited application was the generation of quizzes and assessment questions. Faculty also described using GenAI to develop real-world scenarios for assignments, update or rework existing assignments by incorporating prior materials, and create multiple versions of midterm examinations. 

Faculty further reported using generative AI to improve the quality of their assignments. Several participants noted that generating assignment rubrics with AI was more efficient than creating them manually. Others described using AI to improve the clarity of assignment instructions, which they felt enhanced communication between instructors and students. Some faculty observed a reduction in student clarifying questions as a result of clearer AI assisted instructions. Additionally, participants reported using generative AI to summarize audio feedback into brief written comments for students and to assist with drafting responses in courses with high volumes of student emails.

Generative AI was also used to support instructional content development. Some faculty reported using AI to generate images for course materials that were difficult to locate through traditional search engines, as well as to assist in creating lecture slides. Beyond content creation, participants described using AI as a diagnostic tool to help identify areas where instructional materials or explanations may not be effectively supporting student understanding, enabling targeted revisions.

\textit{Participants also noted several concerns related to the use of GenAI in course design and preparation.} Faculty reported \textit{increased effort} associated with iterative refinement of AI generated content and challenges with formatting assessment questions to meet course standards. Additionally, participants emphasized the need to carefully review and verify AI generated solutions to ensure accuracy before use. 

\subsubsection{\textit{Directly supporting students' learning:}} Faculty reported actively encouraging the use of GenAI tools as part of student learning activities. In many courses, students are guided to use GenAI tools to brainstorm and refine project ideas by starting with a vague concept and iteratively prompting the tool to improve clarity and scope. This approach is commonly applied to project-based work. In CS courses, faculty may encourage students to first generate code snippets using GenAI, then integrate and orchestrate these components into a larger and coherent solution. In chemistry courses, students are asked to use GenAI to produce Python code for data visualization and analysis. Similarly, in engineering courses, students may submit pseudocode as part of their assignments and then use GenAI to help translate that pseudocode into source code of a specific programming language. In computer networking classes, students may leverage GenAI to identify anomalous network packets or to convert wireframes into HTML code.

Additionally, several faculty members described assignments designed to promote critical evaluation of AI generated output. One common strategy requires students to produce two solutions to the same problem, one written entirely by the student and one generated by a GenAI tool, followed by a detailed analysis and reflection comparing the quality, logic, accuracy, and relevance of the two solutions. Other faculty ask students to conduct a standard human-led code review and then request an AI generated review of the same code, requiring students to compare, critique, and assess the strengths and limitations of each perspective. 

\textit{A concern raised by faculty is the challenge students face when working with AI generated code beyond its initial creation.} While GenAI can often produce a solid starting point, students frequently struggle during the debugging phase because they did not develop the original logic and therefore do not fully understand how the code works. Faculty also emphasized that AI generated solutions typically require significant refinement to function correctly and/or to meet specific assignment requirements. Students who rely heavily on GenAI may find it difficult to modify specific components of an algorithm or adapt the code to new constraints if they lack a strong foundational understanding.

\subsubsection{\textit{Assisting with administrative tasks and communications:}} Faculty participants also mentioned that they use GenAI tools for administrative and communicative purposes. For example, GenAI is used to summarize lengthy email threads, draft and respond to student emails. Faculty also used it to condense their audio feedback into a few concise sentences, which helped speed up the grading process. Additionally, faculty observed that when they provided their own raw ideas, GenAI was able to preserve their personal voice, producing feedback that felt authentic while reducing the effort required to refine it. 

At the same time, \textit{faculty raised concerns about the potential overuse of GenAI in communication.} They noted that heavy reliance on AI generated messages may erode trust between students and instructors, as students can perceive a lack of genuine interaction. To mitigate this risk, faculty emphasized the importance of transparency, suggesting that being upfront with students about the use of GenAI in correspondence can help maintain trust.

\subsection{RQ2: What benefits and challenges instructors observe regarding student                           learning? }

Faculty participants reported a range of perceived benefits and challenges associated with the use of GenAI in the classroom, particularly in relation to its impact on student learning.

\subsubsection{\textit{Benefits:}} One of the primary benefits observed, particularly in CS courses, is the way GenAI helps bridge technical gaps for students. Faculty noted that a greater number of students are now able to complete programming assignments, suggesting that GenAI tools help students overcome obstacles that might previously have caused them to stall or fail. Additionally, faculty observed that GenAI enables students to more rapidly develop ideas and implement code, supporting faster iteration and experimentation during the learning process. Even in classes where students may lack technical coding knowledge, such as chemistry, faculty found that GenAI can provide a complete set of code to perform specific tasks like data processing and curve fitting, tasks that would be difficult for students to write from scratch.

Faculty also observed students appeared to complete their work more quickly and were more likely to submit assignments on time, as GenAI helped them move through the initial stages of projects more efficiently. Additionally, faculty noted that students who used GenAI to brainstorm and refine project topics were able to articulate their ideas more clearly and rapidly, which accelerated their progress toward final project implementation.

Faculty also described GenAI functioning like a ``private teaching assistant" for students, offering support outside of class time. Students frequently use GenAI tools to receive immediate responses to homework questions, often turning to these tools before attending office hours or seeking in-person tutoring. This immediate availability was seen as especially beneficial for students balancing coursework with jobs or family responsibilities. Faculty further noted that GenAI helps students overcome learning bottlenecks more quickly than traditional web searches.

\subsubsection{\textit{Challenges:}} Faculty participants identified several challenges related to students’ use of GenAI. One concern is the increased prompting burden placed on students, who may need to spend considerable time crafting precise prompts to obtain usable results. Faculty also pointed out that when students use specialized GenAI tools, the quality of the results can be uneven, meaning much of the generated content still needs to be evaluated, corrected, or discarded. Most concerning, instructors cautioned that over-reliance on GenAI may allow students to bypass critical thinking and problem-solving processes, reducing opportunities for meaningful learning and skill development.

Faculty raised significant concerns about academic integrity and plagiarism related to student use of GenAI. They emphasized the growing difficulty of identifying AI generated work. Instructors noted that attempting to detect misuse can be time-consuming and often detracts from core teaching responsibilities. This challenge is compounded by the unreliability of current GenAI detection tools, which may produce false positives or fail to identify more sophisticated uses. In some disciplines, particularly introductory technical courses, proving misconduct is especially difficult because acceptable solutions are often inherently similar. As a result, some faculty have moved away from attempts to ``AI proof" assignments. They view such efforts as unsustainable and often counterproductive. To uphold academic integrity, many instructors are instead returning to traditional assessment methods. These include in class, pen-and-paper exams, oral exams, and in-person explanations of submitted work.



\subsection{RQ3: What institutional resources and policies are needed to support                        effective generative AI adoption? }

Focus group participants identified several key areas where stronger institutional support is needed. These areas include expanded training and resources for faculty and students, clearer policies and guidelines for GenAI use, and support for meaningful curricular and program-wide changes. 

\subsubsection{\textit{Training and resources:}} One of the primary forms of institutional support identified by faculty was professional development workshops. Faculty emphasized the importance of workshops focusing on the fundamentals of GenAI, particularly how large language models work, noting that they cannot effectively teach with GenAI if they do not understand its underlying mechanisms. Instructors stressed the need to understand that these tools do not \textit{think} or have emotions, but instead generate responses by predicting text based on learned patterns. In addition, faculty expressed interest in workshops on prompt engineering. They highlighted the value of learning how to ask better questions and how to iteratively refine prompts, noting that effective prompting often leads to high quality outputs. Faculty also requested task-specific training, such as workshops on using GenAI to create grading rubrics or generate lecture materials.

Participants also emphasized the importance of having a centralized repository of resources related to GenAI. Faculty expressed interest in a shared collection of reusable prompts that could be adapted for similar instructional tasks. In addition, they highlighted the value of including case studies that document both successful and unsuccessful uses of GenAI. 

Faculty also proposed additional forms of institutional support. One suggestion was the creation of a dedicated consultation service or core AI support team that instructors could approach with specific technical, pedagogical, or ethical questions, rather than relying solely on general IT support. Faculty also mentioned the need for communities of practice, such as \textit{AI Squares}, modeled after existing practice of \textit{teaching squares} at SFSU. Each teaching square is typically constituted of four multi-disciplinary instructors. These groups meet regularly and cultivate a friendly place for peer-to-peer learning, resource sharing, and discussions across disciplines.

Participants also highlighted the need for institutional investment in time and staffing. They emphasized the lack of compensated time to redesign courses and requested paid release time or professional development funds to support curriculum restructuring in response to GenAI. Faculty also called for increased resource allocation to support infrastructure and to hire personnel for dedicated AI support teams.

\subsubsection{\textit{Policy and guidelines:}} Faculty identified a consistency gap in GenAI policies across courses, noting that students are often confused when GenAI use is permitted in some classes but discouraged or penalized in others. They emphasized the need for clear, department-level guidelines so students understand what forms of GenAI use are acceptable without fear of unintended consequences. Faculty also noted the need for coordination between courses. They emphasized that GenAI policies should be aligned across sequential classes through department or college level collaboration to ensure consistent expectations for students. Faculty expressed interest in institutional syllabus templates and clear rules for GenAI use. They noted that such frameworks reduce the burden of writing policies from scratch.

Participants described the current state of intellectual property and citation related to GenAI as unclear and inconsistent. They emphasized the need for institutional agreement on academic honesty in this new context. Participants also noted confusion around how to properly cite AI generated content, as existing guidance often lacks clear citation standards.

Participants also raised broader ethical and equity concerns related to institutional adoption of GenAI. They highlighted accessibility issues, noting that some AI tools create barriers for students with disabilities and may raise compliance concerns that policy must address. Participants also emphasized the need for explicit guidance and training on bias in AI, including how issues related to race, language, and gender can affect student experiences.

\subsubsection{\textit{Curricular and program changes:}} Participants across the focus groups discussed the need for a mandatory, university wide AI literacy course. The goal is to level the playing field for all students. One suggestion is to use upcoming degree restructuring to implement a required first-year seminar. This course would focus on AI literacy, ethics, and acclimation to campus life. It could also count toward multiple General Education requirements. Another perspective emphasizes that AI is now used across all industries. Because of this, a mandatory AI for Everyone style course is needed for all students, not just those in technical fields.

Faculty discussed the need to rethink how major requirements are structured, especially the progression from introductory to advanced courses. Participants noted differences in how GenAI might be treated in lower versus upper division courses, the need for better coordination of GenAI related expectations across prerequisites, and concerns about whether current technical and lab based courses remain relevant given AI’s ability to generate answers easily.

Faculty also discussed the need for greater institutional support to enable meaningful curricular redesign related to AI and GenAI. Participants emphasized the lack of paid time or resources for faculty to adjust teaching and curriculum, the value of a centralized team to guide AI integration across departments, and the importance of investing in staffing, infrastructure, and computing resources to support advanced coursework and research.

At the same time, faculty expressed caution about making permanent institutional changes too quickly. Participants noted the difficulty of predicting how GenAI will reshape higher education in the near future and warned against basing major structural decisions on short term trends. There was also an emphasis on allowing curriculum changes to emerge through department level discussions rather than through broad, top down institutional mandates.

\section{Discussion}


\subsubsection{\textbf{\textit{Shift in faculty labor:}}} Findings from \emph{RQ1} show that while most faculty participants use GenAI for classroom tasks such as quiz generation, this has not necessarily reduced their workload. Instead, faculty report spending less time creating content from scratch and more time reviewing, refining, and verifying AI generated outputs. Thus, GenAI appears to shift faculty labor from content creation to expert curation, rather than eliminating it. This observation raises questions about whether GenAI leads to genuine gains in faculty efficiency.

\subsubsection{\textbf{\textit{Illusion of student competency:}}} Findings from \emph{RQ1} indicate that assignment submission rates are higher when students use GenAI compared to when they do not. However, faculty also report that many students struggle to debug AI generated code due to a lack of fundamental understanding of the underlying concepts. This suggests that GenAI can mask students' underlying incompetencies under current assessment practices. These findings raise important questions about whether existing assessment methods adequately measure student competence and whether new approaches to evaluation are needed.

\subsubsection{\textbf{\textit{Shift in evaluation methods:}}}
Findings from RQ2 indicate that some faculty are reverting to more traditional forms of evaluation, such as in class pen and paper exams and oral assessments, where the use of GenAI is restricted or prohibited. At the same time, some faculty are incorporating GenAI into coursework by designing assignments that require students to compare and contrast AI generated outputs with those produced by humans. This dual approach reflects ongoing efforts to preserve academic integrity while promoting critical engagement with GenAI.
\subsubsection{\textbf{\textit{Needs for critical thinking interventions:}}} Throughout our focus group study, faculty's one constant concern is that students may unintentionally (and uncritically) become over-reliant on GenAI, which in turn can stall or even negatively impact their critical thinking and problem solving skills. Besides investigating various assessment strategies as discussed above, it becomes ever urgent for educators to actively and deliberately introduce pedagogical interventions to ``force" students into the habit of developing their critical thinking skills in spite of the prevalence of GenAI tools. For example, the following strategies have been reported in the literature to be effective: reframing GenAI responses as questions, facilitating conflict-filled group discussions, reflecting on the process, and gamifying a learning task~\cite{lee2025impact}.
\subsubsection{\textbf{\textit{Balancing AI policy and autonomy under uncertainty:}}}
RQ3 underscores the need to balance consistent, university-wide GenAI policy guardrails that reduce student confusion with department-level autonomy that allows programs to use AI in discipline-specific ways. Faculty expressed skepticism toward static mandates and cautioned against implementing permanent institutional changes prematurely, given uncertainty about how GenAI will reshape both higher education and the future workforce. The findings suggest that policy alignment alone is insufficient; it must be accompanied by sustained institutional investments in faculty development, shared pedagogical resources, and a dedicated core AI consultation function.

At the same time, fundamental questions remain open regarding which skills will be most critical for the future workforce, likely varying across industry sectors, and how higher education should redesign curricula to cultivate those skills in an AI-rich environment. Addressing these uncertainties will continue to shape the ongoing evolution of institutional AI guidelines.
\subsubsection{\textbf{\textit{Limitations:}}} We note that the findings are contextually specific to faculty in typical STEM disciplines at a single institution. The number of participants is also relatively small. Consequently, our findings may not apply to non-STEM fields, other institutional types, or geographic regions.

\section{Conclusions and Future Work}

In this work, we conducted a focus group study in a semi-controlled manner with 29 STEM faculty members from the College of Science and Engineering at San Francisco State University. This study provides insights into how STEM faculty are currently engaging with GenAI in teaching and learning, highlighting shifts in instructional practice, emerging challenges in assessment, strong concerns in likely erosion of learning, and the need for balanced institutional support. Our findings suggest that integrating GenAI into higher education extends beyond tool adoption and requires deliberate and critical reconsideration of pedagogical practices, evaluation methods, and governance structures. 

As GenAI technologies continue to evolve, future work will examine longitudinal changes in faculty practices and student learning outcomes, investigate the effectiveness of AI-informed assessment and critical thinking interventions, and explore how institutional policies can remain adaptive across disciplines and workforce contexts. We also plan to incorporate student perspectives and conduct cross-institutional studies to further inform the design of equitable and sustainable approaches to GenAI integration in higher education.

\section{Acknowledgments}
This work was partially supported by the project \emph{E-GAISE: Ethical Generative AI in Undergraduate Computer Science Education}, funded through the AI FAST Challenge Grants program of the California Education Learning Lab.

This study received Institutional Review Board (IRB) exemption approval from San Francisco State University.

\bibliography{aaai2026}

\end{document}